\documentclass[twoside]{article}
\usepackage{fleqn,espcrc2,axodraw}
\usepackage{graphicx}

\newcommand{\AmS}{{\protect\the\textfont2
  A\kern-.1667em\lower.5ex\hbox{M}\kern-.125emS}}

\begin{document}

% declarations for front matter
\title{\vspace*{-10mm} \hfill {\small CTS-IISc-14/00}\\
\vspace*{5mm}
A derivation of Regge trajectories in large$-N$ transverse lattice QCD%
\thanks{Presented at ``Lattice 2000'', 17-22 August 2000, Bangalore, India.
       To appear in the proceedings.}}

\author{Apoorva D. Patel\address{CTS and SERC, Indian Institute of Science,
        Bangalore-560012, India}
        (e-mail: adpatel@cts.iisc.ernet.in)}
       
\begin{abstract}
Large-N QCD is analysed in light-front coordinates with a transverse lattice
at strong coupling. The general formalism can be looked up on as a $d+n$
expansion with a stack of $d-$dimensional hyperplanes uniformly spaced in
$n$ transverse dimensions. It can arise by application of the renormalisation
group transformations only in the transverse directions. At leading order in
strong coupling, the gauge field dynamics reduces to the constraint that only
colour singlet states can jump between the hyperplanes. With $d=2$, $n=2$
and large$-N$, the leading order strong coupling results are simple
renormalisations of those for the 't Hooft model. The meson spectrum lies on
a set of parallel trajectories labeled by spin. This is the first derivation
of the widely anticipated Regge trajectories in a regulated systematic
expansion in QCD.
\end{abstract}

\maketitle

\section{THE CORNERSTONES}

The systematic study of strong interactions started with the discovery of
lots of hadrons and the $S-$matrix analysis of their interactions. An
important ingredient in this analysis was the observation that the hadron
masses, when plotted in the $M^2-J$ plane, lay on a set of approximately
linear trajectories---the Regge trajectories---labeled by various quantum
numbers. This phenomenological feature subsequently led to dual resonance
models and string theory. Later on quark models and QCD replaced all the
bootstrap ideas, but still it is generally accepted that QCD will reproduce
an effective string description of the strong interactions at long distances.

Wilson's introduction of a non-perturbative lattice regulator and strong
coupling expansion provided a qualitative realisation of this expectation
\cite{WILSON}. In brief \cite{CREUTZ}:\\
$\bullet$ Area law for a large Wilson loop demonstrates linear confinement.\\
$\bullet$ In the strong coupling limit, $g\rightarrow\infty$, the gauge
field action vanishes and the gauge field becomes completely random.\\
$\bullet$ The strong coupling expansion in inverse powers of $g$ has a
non-zero radius of convergence. It is the analogue of the high temperature
expansion used in statistical mechanics.\\
$\bullet$ The strong coupling limit can be understood as the extreme stage
of renormalisation group evolution. The regulating cutoff is lowered as
much as possible. All the excited states are integrated out, leaving behind
only the lowest state in each quantum number sector of the theory.\\
$\bullet$ As $g\rightarrow\infty$, the gauge field dynamics is reduced to
the constraint that at every space-time point all the fields must combine
into a gauge singlet.\\
$\bullet$ The strong coupling limit is not universal, and results depend on
the type of lattice discretisation used. In any case, only subgroups of the
continuum Lorentz and chiral symmetries remain exact.\\
$\bullet$ As $g\rightarrow\infty$, the string tension diverges and glueballs
disappear from the theory. But hadrons containing quarks retain finite
mass.\\
$\bullet$ Despite non-universality, the lightest hadron masses follow a
pattern qualitatively in agreement with their experimental values.
Quantitative fits can be obtained with about 30\% accuracy.\\
$\bullet$ With only one surviving state in each quantum number sector,
form factors show pole dominance. In coordinate space, the tails are
exponential (and not gaussian as is often assumed in phenomenological
analyses).\\
$\bullet$ The strong coupling analysis can be carried out with any gauge
group in any number of dimensions. In 4-dimensional QCD, numerical
simulations show that the strong coupling region is analytically connected
to the weak coupling region.

't Hooft introduced the expansion in inverse powers of the number of colours
as another method of simplifying the non-abelian gauge field dynamics.
The method could obtain the meson spectrum in 2-dimensions for any value
of the gauge coupling \cite{THOOFT}. The important features of this solution
are \cite{COLEMAN}:\\
$\circ$ The large$-N$ limit has only planar diagrams.\\
$\circ$ Choosing an axial gauge in two dimensions eliminates the non-linear
self-coupling of the gauge fields.\\
$\circ$ The dynamics of the gauge field is replaced by a linearly confining
potential.\\
$\circ$ The ultraviolet regulator can be removed, while the infrared
divergence is controlled by the principle value prescription.\\
$\circ$ Use of light-front coordinates simplifies the Lorentz index
structure of the theory and eliminates non-dynamical variables, making
Lorentz invariance manifest.\\
$\circ$ The light-front wavefunctions are just the parton structure
functions appropriate for analysis of deep inelastic scattering.\\
$\circ$ Spin does not exist in the 2-dimensional theory, but parity is
a good quantum number.\\
$\circ$ The meson spectrum is determined by the eigenvalues of an integral
equation. It lies on an approximately linear trajectory in the $M^2-n$
plane, where $n$ is the radial excitation quantum number.

In the following sections, a framework is constructed that combines these
two approaches to simplifying the gauge field dynamics. The result is an
analytically tractable systematic expansion in QCD which is closer to
observed phenomenology than any previous attempt.

\section{INCREASING DIMENSIONS\\ OF A THEORY}

When exact results for a $d-$dimensional theory are known, they can be used
to make predictions for the same theory in neighbouring number of dimensions.
A well-known method used with weak coupling perturbation theory is the
$d+\epsilon$ expansion, where physical properties are evaluated as asymptotic
series in the continuous variable $\epsilon$. A method that uses strong
coupling expansions can be constructed too, introducing the explicit
regulator as a transverse lattice. The space-time geometry can be thought
of as a stack of $d-$dimensional hyperplanes, uniformly spaced along $n$
transverse dimensions. Such a geometry is illustrated in Fig.1. It can arise
from the $(d+n)-$dimensional theory by anisotropic renormalisation group
evolution, where the cutoff is lowered only in $n$ transverse dimensions.

\begin{picture}(250,200)(0,0)
\Line(0,0)(0,150) \Line(0,0)(40,15)
\Line(0,150)(80,180) \Line(80,165)(80,180)
\Line(40,0)(40,150) \Line(40,0)(80,15)
\Line(40,150)(120,180) \Line(120,165)(120,180)
\Line(80,0)(80,150) \Line(80,0)(120,15)
\Line(80,150)(160,180) \Line(160,165)(160,180)
\Line(120,0)(120,150) \Line(120,0)(200,30)
\Line(120,150)(200,180) \Line(200,30)(200,180)
\Text(60,190)[c]{.} \Text(65,190)[c]{.} \Text(70,190)[c]{.} \Text(75,190)[c]{.}
\Text(205,190)[c]{.} \Text(210,190)[c]{.}
\LongArrow(95,190)(80,190) \Text(100,190)[c]{$a_\perp$} \LongArrow(105,190)(120,190)
\LongArrow(135,190)(120,190) \Text(140,190)[c]{$a_\perp$} \LongArrow(145,190)(160,190)
\LongArrow(175,190)(160,190) \Text(180,190)[c]{$a_\perp$} \LongArrow(185,190)(200,190)
\Vertex(20,30){3} \Text(20,20)[c]{$i$}
\Vertex(170,150){3} \Text(170,160)[c]{$f$}
\ArrowArc(190,115)(40,120,230) \ArrowArc(150,115)(40,300,60)
\Photon(152,125)(188,125){3}{5} \Photon(152,105)(188,105){3}{5}
\DashLine(125,80)(170,80){5} \DashLine(125,85)(165,85){5}
\ArrowArc(108,90)(20,210,330) \ArrowArc(108,75)(20,30,150)
\Photon(108,70)(108,95){3}{4}
\DashLine(60,80)(91,80){5} \DashLine(60,85)(91,85){5}
\ArrowArc(75,105)(30,124,236) \ArrowArc(45,110)(30,304,56)
\Photon(45,110)(75,105){3}{5}
\DashLine(23,130)(60,130){5} \DashLine(21,135)(60,135){5}
\ArrowArc(-53,80)(90,323,34) \ArrowArc(93,80)(90,143,217)
\Photon(6,60)(34,60){3}{5} \Photon(6,100)(34,100){3}{5}
\end{picture}

\medskip
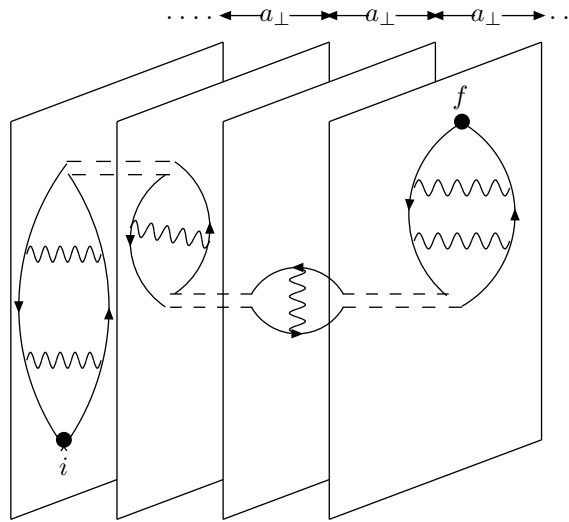
\noindent Figure 1. A schematic representation of the meson propagator on a
transverse lattice at strong coupling. The jumps between hyperplanes are
shown by dashed lines.
\medskip

When the interaction between the hyperplanes becomes weak, we can use the
exact results of the $d-$dimensional theory, to find the properties of the
anisotropic $(d+n)$ dimensional theory. The expansion we need then is an
expansion in the number of interactions between the hyperplanes\footnote{%
The subscript $\perp$ will be used to denote variables corresponding to
the transverse directions.}.
In models of statistical mechanics, as $T_\perp\rightarrow\infty$, the
hyperplanes decouple and the theory reduces to multiple copies of the
$d-$dimensional theory. In gauge theories, as $g_\perp\rightarrow\infty$,
charged states cannot propagate between hyperplanes, but neutral states
can (see Fig.1). The resultant theory is then not the $d-$dimensional
gauge theory, but related to it. Provided we can work out this relation,
we can use the exact properties of the $d-$dimensinal theory to predict
the properties of the $(d+n)-$dimensional theory.

Conventionally, strong coupling (or high temperature) expansions have been
constructed treating all directions on an equal footing (i.e. $d=0$).
With a non-trivial start using the $d\ne0$ exact solution, it is
reasonable to expect that the anisotropic expansion would provide a better
idea of the weak coupling behaviour of the $(d+n)-$dimensional theory.
It is important to note that this strategy can be used only in situations
when the $d-$ and the $(d+n)-$dimensional theory are in the same phase.
(For example, the connection between the $2-$dimensional Schwinger model
to $4-$dimensional QED is established in the strong coupling phase with a
massive photon, and not in the weak coupling phase with a massless photon.)

The formalism of $(d+n)-$expansion is general enough to be applied to many
other physical problems, e.g., layered high-temperature superconductors
where Cooper pairs can hop between layers but individual electrons cannot,
or higher-dimensional extensions of the standard model where the extra
dimensions are accessible to only certain neutral particles. Here we use
it to connect 4-dimensional large$-N$ QCD to the 2-dimensional 't Hooft
model.

\section{STRONG COUPLING TRANSVERSE LATTICE LARGE$-N$ QCD}

With a transverse lattice, the coordinate space of QCD is reduced from
$R^4$ to $R^2 \times Z^2$. It can also be thought of compactifying the
momentum space of QCD to $R^2 \times T^2$. Transverse lattice formulation
of QCD was introduced by Bardeen and Pearson long back \cite{BARDEEN}.
Its strong coupling limit, however, has not been thoroughly investigated.

The transverse gauge field components are represented by the unitary link
matrices, $U_\perp(x) = \exp(ia_\perp A_\perp(x))$, while the hyperplane
gauge field components are the usual $A_\mu$. Taking the strong coupling
limit in the transverse directions replaces the dynamics of $U_\perp$ to
constraints, i.e., only states contracted to colour singlets can propagate
in the transverse direction. Choosing the light-front axial gauge $A^+=0$,
the hyperplane gauge field is reduced to linear confining potential as in
the 't Hooft model. Thus all the gauge dynamics is simplified, and it is
easy to study the bound states of quarks\footnote{%
Numerical simulations of transverse lattice QCD have mostly concentrated
on the pure gauge sector \cite{DALLEY}, which is more complicated in the
strong coupling limit than the quark sector.}.
The $\gamma-$matrices are not eliminated. But they are essential for
defining the spin, and they can be handled in the same manner as in
conventional strong coupling QCD.

The parameters of QCD are the gauge coupling and the quark masses.
In lattice calculations, they are often traded off for the lattice
cutoff and the hopping parameters, respectively. The anisotropic
geometry increases the number of parameters; it is convenient to
choose them as $m$ and $g$ for the $2-$dimensional hyperplanes and
$a_\perp$ as the transverse lattice spacing. Thus the action can be
written as ($g$ is held fixed as $N\rightarrow\infty$)
\begin{eqnarray}
S &=& {a_\perp^2 N \over g^2}\sum_{x_\perp}\int d^2x~
      \Big[ -{1 \over 4} F_{\mu\nu b}^a(x) F_a^{\mu\nu b}(x) \nonumber\\
  &+& \overline{\psi}(x) (i\gamma^\mu \partial_\mu
                         - \gamma^\mu A_\mu - m) \psi(x) \nonumber\\
  &+& {i \over 2a_\perp} \big(
      \overline{\psi}(x) \gamma^\perp U_\perp(x) \psi(x+a_\perp) \nonumber\\
  &-& \overline{\psi}(x) \gamma^\perp U_\perp^\dag(x-a_\perp)
                                      \psi(x-a_\perp) \big) \Big] ~~,
\end{eqnarray}
together with the constraint that in any correlation function products of
$U_\perp(x)$ must contract to a colour singlet at each space-time point.
The functional form of the lattice discretisation is not unique, and here
I have chosen the naive fermion prescription for simplicity.

A mixed representation is suitable for explicit calculations---momentum
space representation for the hyperplane variables (so they can be dealt
with in the same way as in the 't Hooft model) and coordinate space
representation for the transverse variables (so their colour singlet
constraint can be enforced easily). For example, the hyperplane gauge
field propagator becomes
\begin{equation}
D_{\mu\nu}(k;x_\perp,x'_\perp) = i \delta_{\mu+}\delta_{\nu+}
  \delta_{x_\perp,x'_\perp} {\rm P}{1 \over (k_-)^2} ~~.
\end{equation}
The transverse lattice spacing, $a_\perp$ converts the dimensionless
QCD coupling $g$ to the dimensionful coupling of the 't Hooft model.

\begin{picture}(250,120)(0,-10)
\Text(20,-15)[c]{$(a)$}
\Vertex(10,0){2} \Text(25,0)[c]{$i$}
\Vertex(10,100){2} \Text(25,105)[c]{$f$}
\CArc(10,20)(20,270,0) \CArc(10,20)(20,35,85) \CArc(10,20)(20,95,270)
\Line(8,40)(8,60) \Line(12,40)(12,60)
\CArc(10,80)(20,275,325) \CArc(10,80)(20,0,90) \CArc(10,80)(20,90,265)
\Line(26,30)(26,70) \Line(30,20)(30,80)
\Text(90,-15)[c]{$(b)$}
\Vertex(80,0){2} \Text(95,0)[c]{$i$}
\CArc(80,20)(20,270,85) \CArc(80,20)(20,95,270)
\Line(78,40)(78,60) \Line(82,40)(82,60)
\CArc(80,80)(20,275,90) \CArc(80,80)(20,90,265)
\Vertex(80,100){2} \Text(95,105)[c]{$f$}
\PhotonArc(80,50)(30,310,50){2}{7}
\Text(160,-15)[c]{$(c)$}
\Vertex(150,0){2} \Text(165,0)[c]{$i$}
\CArc(150,20)(20,270,85) \CArc(150,20)(20,95,270)
\Line(148,40)(148,60) \Line(152,40)(152,60)
\CArc(150,80)(20,275,90) \CArc(150,80)(20,90,265)
\Vertex(150,100){2} \Text(165,105)[c]{$f$}
\PhotonArc(150,50)(35,305,55){2}{7}
\PhotonArc(150,50)(29,310,50){2}{7}
\end{picture}

\bigskip 
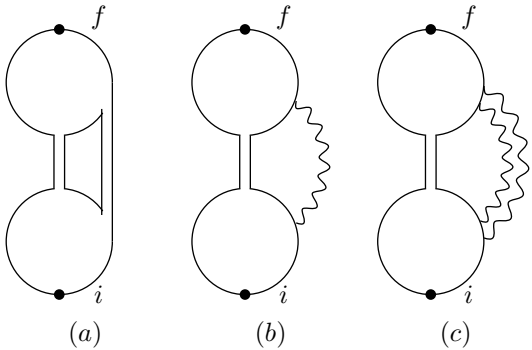
\noindent Figure 2. Corrections to the strong coupling transverse lattice
large$-N$ QCD: (a) a quark loop is suppressed by $N^{-1}$, (b) a gluon jump
between hyperplanes is suppressed by $N^{-1}g_\perp^{-2}$, (c) a glueball
jump is suppressed by $N^{-2}$.
\medskip

As illustrated in Fig.1, while propagating along the hyperplanes, the
quarks spread out and exchange gluons. But for $g_\perp\rightarrow\infty$,
they must come together and form colour singlets while jumping from one
hyperplane to another. The corrections to this leading behaviour arise
from multiple connections amongst parts of propagators on different
hyperplanes, as shown in Fig.2. All these corrections involve loops and
are suppressed at least by $N^{-1}$. The leading behaviour of the hadron
propagators is thus ``tree-like''.

\section{THE MESON SPECTRUM}

It is easy to work out the explicit formulae for the propagators. The
quark propagator receives gluon renormalisations as in the 't Hooft model,
but it also gets corrections from hops back and forth between the
hyperplanes (see Fig.3a). These additional corrections can be calculated
in a closed form. It is sufficient to note, however, that they are tadpoles
and cannot carry away any momentum or colour. Their only effect is to 
renormalise the quark mass, and since quark masses are ultimately fit
parameters to reproduce the hadron masses, we do not need the explicit
form for the tadpoles. Let $\tilde{m}$ and $\tilde{\kappa}_\perp$ be the
renormalised mass and hopping parameters, after including all tadpole
corrections. It is worthwhile to observe that the same tadpoles
contribute to the chiral condensate $\langle\overline{\psi}\psi\rangle$,
and so this mass renormalisation can be understood as generation of the
constituent quark mass due to hops in the extra transverse directions.

\begin{picture}(250,110)(0,0)
\Text(40,10)[c]{$(a)$}
\ArrowLine(0,40)(40,40) \ArrowLine(44,40)(84,40)
\PhotonArc(42,40)(16,180,0){2}{6}
\Line(40,40)(40,60) \Line(44,40)(44,60)
\CArc(42,80)(20,275,265) \Photon(22,80)(62,80){2}{6}
\Text(150,10)[c]{$(b)$}
\Vertex(110,60){2} \Text(100,60)[c]{$\Gamma_i$}
\Vertex(190,60){2} \Text(200,60)[c]{$\Gamma_f$}
\CArc(126,62)(16,0,180) \CArc(126,58)(16,180,0)
\Line(142,58)(158,58) \Line(142,62)(158,62)
\CArc(174,62)(16,0,180) \CArc(174,58)(16,180,0)
\Photon(126,42)(126,78){2}{5} \Photon(174,42)(174,78){2}{5}
\end{picture}

\smallskip
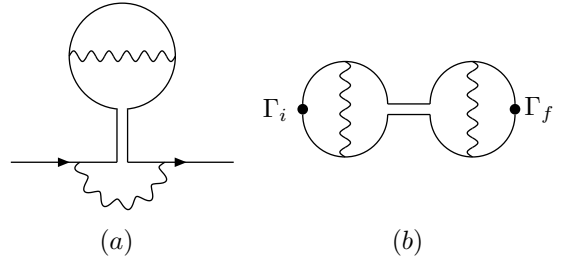
\noindent Figure 3. Modifications of the 't Hooft model results due to
transverse lattice at strong coupling: (a) tadpole correction to the quark
propagator, (b) jump correction to the meson propagator.
\medskip

In order to evaluate the meson propagators, we have to keep track of the
$\gamma-$matrices, unlike the 't Hooft model case. $\gamma-$matrices
appear in the quark propagators as well as in the operators creating
and destroying mesons (see Fig.3b). The anticommutation relations of
$\gamma-$matrices are sufficient to evaluate the spinor traces, and
the techniques for their evaluations are well-known in conventional
strong coupling expansions.

\smallskip
\begin{picture}(250,60)(0,0)
\Text(50,0)[c]{$(a)$}
\Vertex(10,10){2} \Text(0,10)[c]{$i$}
\Line(10,8)(32,8) \Line(10,12)(28,12)
\Line(28,12)(28,32) \Line(32,8)(32,28)
\Line(32,28)(52,28) \Line(28,32)(48,32)
\Line(48,32)(48,52) \Line(52,28)(52,48)
\Line(52,48)(68,48) \Line(48,52)(72,52)
\Line(72,52)(72,32) \Line(68,48)(68,28)
\Line(68,28)(90,28) \Line(72,32)(90,32)
\Vertex(90,30){2} \Text(100,30)[c]{$f$}
\Text(150,0)[c]{$(b)$}
\Vertex(110,10){2} \Text(100,10)[c]{$i$}
\Line(110,8)(132,8) \Line(110,12)(128,12)
\CArc(134,22)(12,120,240) \CArc(126,18)(12,300,60)
\Photon(122,20)(138,20){2}{3}
\Line(132,28)(152,28) \Line(128,32)(148,32)
\CArc(154,42)(12,120,240) \CArc(146,38)(12,300,60)
\Photon(142,40)(158,40){2}{3}
\Line(152,48)(168,48) \Line(148,52)(172,52)
\CArc(174,38)(12,120,240) \CArc(166,42)(12,300,60)
\Photon(162,40)(178,40){2}{3}
\Line(168,28)(190,28) \Line(172,32)(190,32)
\Vertex(190,30){2} \Text(200,30)[c]{$f$}
\end{picture}

\medskip
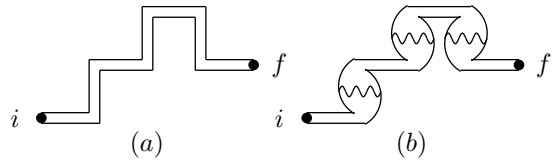
\noindent Figure 4. The strong coupling behaviour of the meson propagator,
when (a) all the dimensions are latticised, (b) only the transverse
dimensions are latticised.
\medskip

As shown in Fig.4, the structure of the meson propagator in the transverse
lattice QCD does not differ very much from that in the conventional strong
coupling QCD. The graphs are ``random walks'' and can be conveniently summed
using generating function techniques. The final result is simple: the complete
inverse meson propagator is the sum of inverse meson propagators corresponding
to propagation in individual directions. Compared to the 't Hooft model result,
the poles of the meson propagator are thus translated due to propagation in
the transverse directions.

It is known from conventional strong coupling expansions that this shift in
the pole position depends on the $\gamma-$matrices used to create and destroy
the meson, and is largely independent of the quark mass. Its explicit form
depends on the lattice discretisation used for fermions in the transverse
direction. For example, for naive fermions, the exact identity
\begin{eqnarray}
&& G(x_i,x_f;\Gamma_i=\Gamma_f=\gamma_\perp) \nonumber\\
&& = (-1)^{x_f-x_i} G(x_i,x_f;\Gamma_i=\Gamma_f=\gamma_5) 
\end{eqnarray}
fixes the vector meson propagator to be the same as the pseudoscalar meson
propagator shifted in momentum by $\pi/a_\perp$. The net result is that the
meson spectrum lies on a set of parallel trajectories labeled by the
$\gamma-$matrix used to create and destroy the meson, i.e. the spin.
The separation between trajectories depends on $a_\perp$, which can
therefore be determined using the experimental value as an input.

The fact that the meson spectrum consists of a tower of states in each
quantum number channel is not unexpected. A successful combination of
strong coupling and 't Hooft model results is bound to produce such a
spectrum. The surprising part is that the results, though extracted in
an extreme limit of QCD, fit the experimental data remarkably well.
For example, $M_V^2 - M_P^2 \approx const.$, is observed to hold all
the way from $\rho-\pi$ to $B^*-B$.

Thus we have arrived at the elusive Regge trajectories (and their
daughthers) in a straightforward framework based on QCD. Because of
the inherent transverse lattice structure, the theory has only discrete
rotational symmetries. Consequently, the trajectories are labeled by the
spin and not the total angular momentum $J$. Nevertheless, this is the
closest we have got to the real world of strong interactions in a long time.
A more rotationally symmetric regulator (instead of the transverse lattice),
higher order calculations and/or numerical simulations should take us even
closer to the real world.

\section{FUTURE DIRECTIONS}

The above results indicate that strong coupling transverse lattice large$-N$
QCD will be a highly useful phenomenological description of QCD. After all,
it is QCD in an analytically tractable limit, and not an adhoc model. Some
future investigations can be easily pointed out \cite{PATEL}:\\
$\bullet$ It is always possible to choose the reference frame such that the
external legs of $2-$point and $3-$point correlation functions lie on a
hyperplane. Such a choice will facilitate computation of many form factors,
deep inelastic scattering amplitudes and decay matrix elements.\\
$\bullet$ Parameters of effective theories of the strong interactions, such
as the chiral perturbation theory and the heavy quark effective theory, can
be calculated without additional assumptions.\\
$\bullet$ With the light-front coordinates, the calculations automatically
incorporate the Minkowski metric. So strong interaction phase-shifts are in
principle calculable.\\
$\bullet$ The leading contribution of the sea quarks arises from loop graphs
suppressed by $N^{-1}$ (e.g., Fig.2a). This is calculable at the next order
without giving up the $g_\perp\rightarrow\infty$ limit.\\
$\bullet$ Baryons can be included as solitons in this large$-N$ framework.\\
$\bullet$ As mentioned before, the $(d+n)-$expansion can be used to extend
results of many exactly solved low-dimensional theories to higher dimensions.

\end{document}